\documentclass[conference]{IEEEtran}
\IEEEoverridecommandlockouts
\usepackage{cite}
\usepackage{amsmath,amssymb,amsfonts}
\usepackage{algorithmic}
\usepackage{graphicx}
\usepackage{textcomp}
\usepackage{xcolor}
\def\BibTeX{{\rm B\kern-.05em{\sc i\kern-.025em b}\kern-.08em
    T\kern-.1667em\lower.7ex\hbox{E}\kern-.125emX}}
\begin{document}

\title{A Baseline Approach for Modeling and Characterization of Commercial Off-The-Shelf (COTS) Droop Controlled Converter \\

\thanks{This work was supported in part by the Advanced Research Projects Agency-Energy (ARPA-E), U.S. Department of Energy, under Award DE-AR0001580.}
}

\author{
  \IEEEauthorblockN{Muhammad Anees}
  \IEEEauthorblockA{\textit{FREEDM System Center} \\
    \textit{North Carolina State University}\\
    Raleigh, USA \\
    manees@ncsu.edu}
\and
        \IEEEauthorblockN{Lisa Qi}
  \IEEEauthorblockA{\textit{ABB Corporate Research Center} \\
    \textit{ABB Inc.}\\
    Raleigh, USA \\
    lisa.qi@us.abb.com}\\

  \IEEEauthorblockN{Srdjan Lukic}
  \IEEEauthorblockA{\textit{FREEDM System Center} \\
    \textit{North Carolina State University}\\
    Raleigh, USA \\
    smlukic@ncsu.edu}   
\and  
   \IEEEauthorblockN{Mehnaz Khan}
  \IEEEauthorblockA{\textit{ABB Corporate Research Center} \\
    \textit{ABB Inc.}\\
    Raleigh, USA \\
    mehnaz.khan@us.abb.com}
}

\maketitle

\begin{abstract}
Due to advancements in power electronics, new converter topologies are introduced day by day. It's hard to get an equivalent model from any manufacturer of any Commercial Off-The-Shelf (COTS) power electronics converters because of intellectual property (IP) and safety concerns. Most COTS products don't reveal the exact topology of the converter as well as the control architecture and corresponding control gains. Because of these reasons, it's very hard to implement an exact equivalent model of COTS converters. Hierarchical control is widely used in microgrid applications, where different control layers are decoupled on the timescale. This article gives a baseline approach to build an equivalent model of any droop controlled COTS converter to mimic its control performance. Bandwidths of the different control loops are estimated along with the control logic from experiments on COTS converter and then hierarchical control is followed to come up with the equivalent model. The approach is verified on COTS CE+T Stabiliti 30C3 converter with its equivalent model build in MATLAB Simulink.
\end{abstract}


\section{Introduction and Literature Review}

DC Microgrids are getting more and more applications due to the advancements in power electronics, simple control - no need for synchronization, inherent DC nature of  resources i.e Photovoltaics and Batteries [1]. Droop is widely used decentralized control approach to ensure uniform power sharing between multiple resources in a DC microgrid [1]. Commonly used linear droop control approaches are classified as power based droop (PV, VP) and current based droop (VI, IV) [2,3]. Adaptive and non linear droops are also investigated by different researchers [4-7] to improve the voltage regulation along with uniform power sharing. 

Linear droop (VI, VP, IV, PV) is a simple proportional control and does not involve dynamics,
any change in output load power/current, will appear directly in the output voltages following the droop curve. To deal with the fluctuations, concept of virtual inertia is borrowed from AC microgrid control [8]. There are two main approaches to implement virtual inertia in DC systems 1) Low Pass Filter (LPF) [8], 2) virtual machine emulation [9,10]. Both approaches are proved equivalent and can be designed to have similar performance by the correct selection of parameters [11]. 

Hierarchical control is used to ensure coordinated power flow from low to high levels in microgrids [5], where tracking control is the faster control that ensures tracking of voltage and current respectively. While designing the control loops for any converter, bandwidth decoupling of control loops is important from control hierarchy perspective. Idea of bandwidth decoupling in standard droop controlled DC systems can be seen in Fig 1. The fastest control loop is usually kept slower than the switching frequency, so that controller don’t get distracted by the switching dynamics. Following the similar hierarchical approach, any upper control is slower than the inner control layer.

\begin{figure}[htbp]
\centerline{\includegraphics{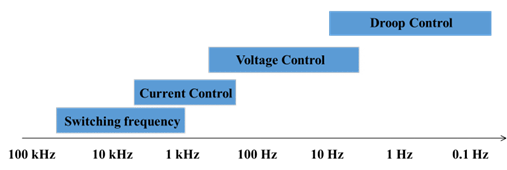}}
\caption{Bandwidth decoupling of droop controlled converter}
\label{fig}
\end{figure}

Most manufactureres don't reveal their converter topology, control structure and corresponding control gains. Considering that it is impossible to come up with an exact model of COTS converter. One approach is to get the impedance scans of the COTS converter and then a small signal model can be estimated [12,13], which can be used for different types of passivity based control [14,15] and EMC analysis [13]. But it is hard to identify the overall control topology and bandwidths of the corresponding control loops just from the impedance scans. 

The main contribution of this paper is to give a baseline approach to identify the bandwidths (BWs) of control loops of a COTS converter and then identify the control logic and come up with an equivalent model. Such equivalent model of an actual COTS product can help the researchers to study the response in a cost effective way for different scenarios including fault, load changes, instability due to constant power loads and current limitations etc. \textbf{CE+T Stabiliti 30C3} converter is chosen as COTS converter for modeling and characterization of control. 

Rest of the paper is organized as follow, section II describes the modeling approach, section III verifies the modeling by comparing the BW of control loops and control logic in equivalent model and COTS converter, Section IV proves the effectiveness of the model through two use cases 1) performance under different droop coefficient/resistance and 2) Instability in DC microgrids due to Constant Power Loads (CPL). At end, section V concludes the article. 
\section{Modeling Approach}
 The chosen COTS converter is a multi port (1x AC, 2x DC) converter, which can be controlled in different ways. As we are interested in knowing the droop behavior so its modeling is simplified for the case when COTS converter is used as droop controlled DC bus. In this study, its other DC port is ignored and input AC side dynamics are also ignored. A DC/DC boost converter with dual loop control regulating with nominal voltage regulation of 350V is chosen as an equivalent switching model of CE+T converter, parameters are given in Table 1. We can choose any other topology of switching converter for equivalent model because switching and converter dynamics can be ignored for the characterization of control behavior of CE+T converter by assuming that fastest control loop is decoupled from switching harmonics (even average model is enough for control design and evaluation). Individual design of control loops is as follow and bode plots are used to verify the control design. 

\begin{table}[htbp]
\caption{Boost Converter Parameters}
\begin{center}
\begin{tabular}{|c|c|}
\hline
\textbf{\textit{Parameters}} & \textbf{\textit{Value}} \\
\hline
Global no-load voltage ref [Vo] & 350V \\
\hline
Source Voltages [E] & 130V \\
\hline
Source resistance [rBAT] & 0.03$\Omega$ \\
\hline
Inductance [L] & 2mH \\
\hline
Inductor ESR [r] & 0.01$\Omega$ \\
\hline
Capacitance [C] & 3.3mF \\
\hline
VP Droop coefficient [Kvp] & 10V/3600W \\
\hline
Equivalent VI droop resistance [K] & 1$\Omega$ \\
\hline
Current control bandwidth & 20KHz \\
\hline
Voltage ramp rate & 50V/sec \\
\hline
Voltage control bandwidth & 200Hz \\
\hline
Low Pass Filter (LPF) bandwidth & 200Hz \\
\hline
\end{tabular}
\label{tab1}
\end{center}
\end{table}

\subsection{Modeling of Current Control Loop}
For the inner current loop  design, transfer function from inductor current (\textit{iL}) to duty cycle (\textit{d}) (Fig. 2) can be formulated using perturbation and linearization around steady state. 
\begin{equation}
    Gid(s)=\frac{((1-Dc) V_o-sLI_L)} {(LCs^2+L/R s+(1-Dc)^2 )} 
\end{equation}

\textit{Dc} is duty cycle and \textit{Rs} is sum of inductor ESR and source resistance. Using (1), as the plant model, bode plot method on loop gain as in (2)  is used to come up with PI gains of the current controller. Bode plot for open loop gain of current control loop is shown in Fig. 3, where we can see that chosen current loop gains results in Phase Margin (PM) of \textit{90 degrees} and Bandwidth of 20\textit{KHz}. \textit{20KHz} bandwidth for current controller is selected in model based on estimation from actual CE+T converter as in subsection A of section III. 
\begin{figure}[htbp]
\centerline{\includegraphics{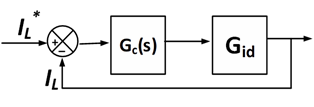}}
\caption{Modeling of current loop}
\label{fig}
\end{figure}
\begin{equation}
    T_c(s)=Gid (s) G_c (s)
\end{equation}
\begin{figure}[htbp]
\centerline{\includegraphics[scale=0.63]{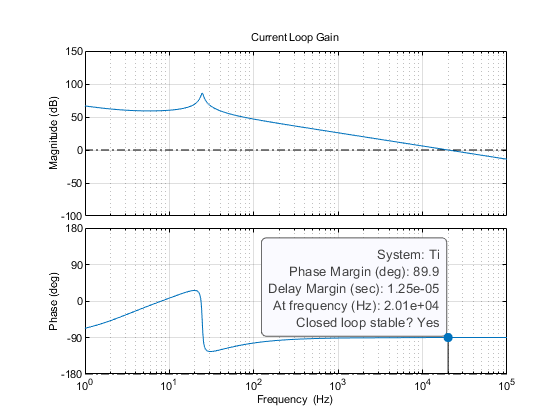}}
\caption{Bode plot for current loop gain}
\label{fig}
\end{figure}
\subsection{Modeling of Voltage Control Loop }
Similar approach is followed for the design of outer voltage loop. Close Loop Transfer Function (CLTF) of inner current loop in (3) is used to calculate the open loop gain \textit{Tv(s) }for outer voltage loop , shown in Fig 4,where (4) represents the plant transfer function for output voltage to inductor current for boost converter. 
\begin{figure}[htbp]
\centerline{\includegraphics[scale=1]{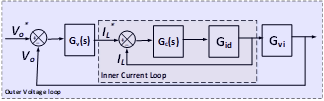}}
\caption{Modeling of voltage loop}
\label{fig}
\end{figure}
\begin{equation}
    Tci (s)=\frac{Gid (s) G_c (s)} {1+Gid (s) G_c (s)} 
\end{equation}
\begin{equation}
    Gvi(s)=\frac{v_o}{i_L} =\frac{(1-Dc) V_o-(sL) I_L} {V_o Cs+2(1-Dc)I_L} 
\end{equation}
\begin{equation}
    T_v(s)=G_vi (s) G_v (s) Tci (s) 
\end{equation}

From some simple tests on COTS CE+T converter, it is found that CE+T has ramp rate controller from \textit{0.25V/sec} to \textit{1000V/sec}. Keeping that in mind a voltage ramp rate controller is implemented on a fast \textit{200Hz} voltage loop.Bode plot design for \textit{200Hz }voltage loop is shown in Fig 5. Ramp rate controller is activated when voltage setpoint is changed, so it is implemented in series with voltage reference command as in Fig 7. 
\begin{figure}[htbp]
\centerline{\includegraphics[scale=0.65]{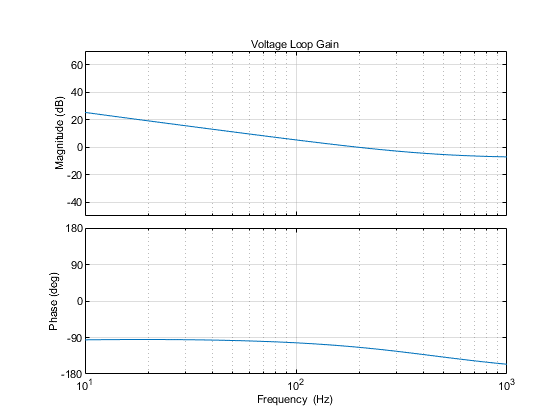}}
\caption{Bode plot for voltage loop gain}
\label{fig}
\end{figure}
\subsection{Modeling of Droop Characteristics}
As confirmed from the conveter's operating manual that CE+T allows maximum droop coefficient upto \textit{1000V/60KW} using VP droop. Following simple test is designed to verify the steady state VP droop characteristics of CE+T for two different droop coefficients as given in Fig. 6. Orange line corresponds to droop coefficient of \textit{500V/60KW }and blue line corresponds to \textit{1000V/60KW} droop coefficient. Droop characteristics on COTS converter are evaluated for steady state no voltage reference of \textit{100V }and power transfer upto \textit{320W} against two mentioned droop gains. Purpose of this test is to evaluate the droop characteristics on COTS converter and then model and design the droop characteristics in equivalent model accordingly. This tests also confirms that droop follows Low Pass Filter (LPF) of about \textit{200Hz}, this will be explained later in the subsections B, section III. LPF is implemented in series with droop cofficient, finalized control scheme is shown in Fig 7. 
\begin{figure}[htbp]
\centerline{\includegraphics[scale=0.60]{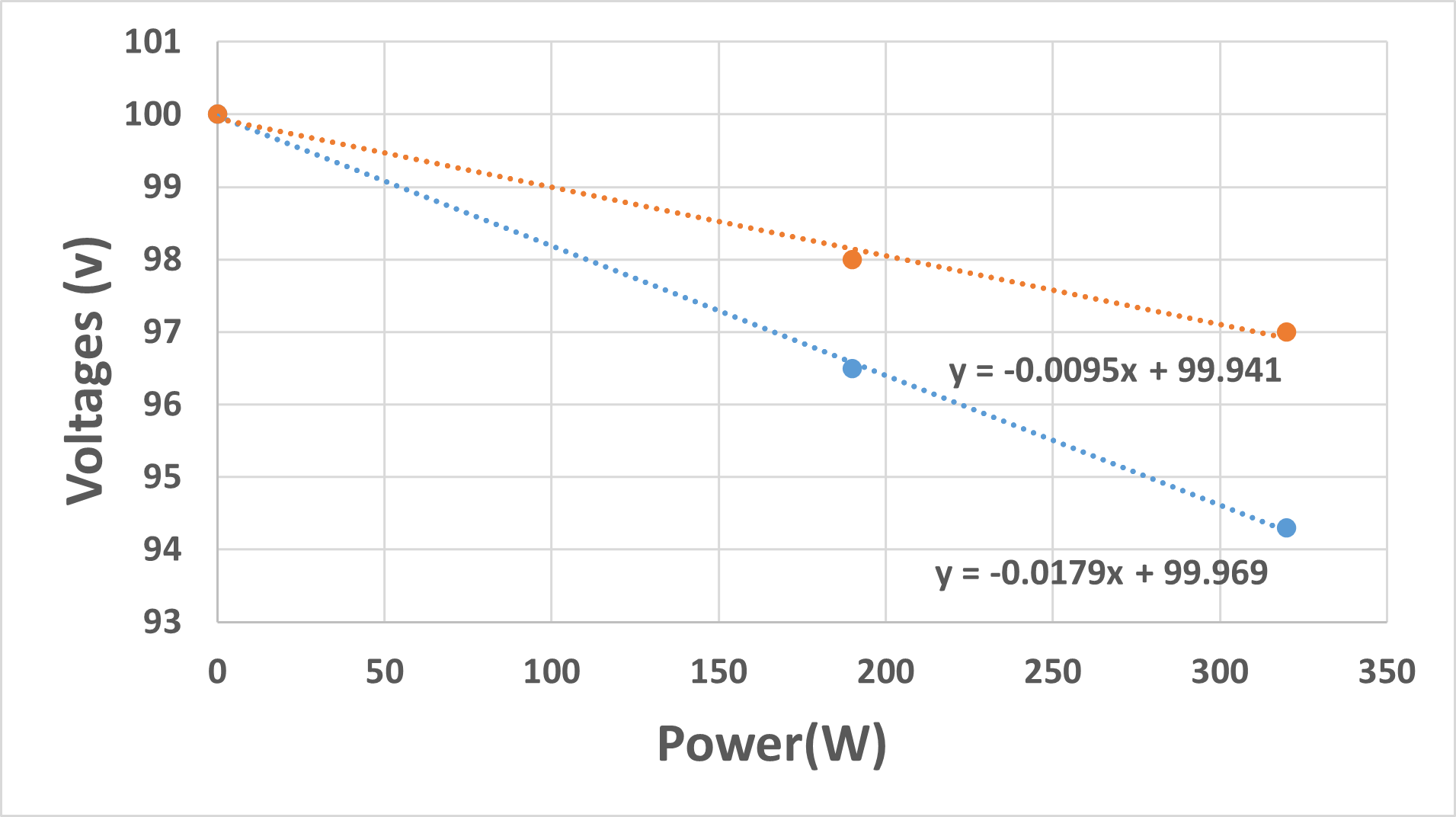}}
\caption{VP droop steady state characteristics for CE+T}
\label{fig}
\end{figure}
\begin{figure}[htbp]
\centerline{\includegraphics[scale=0.65]{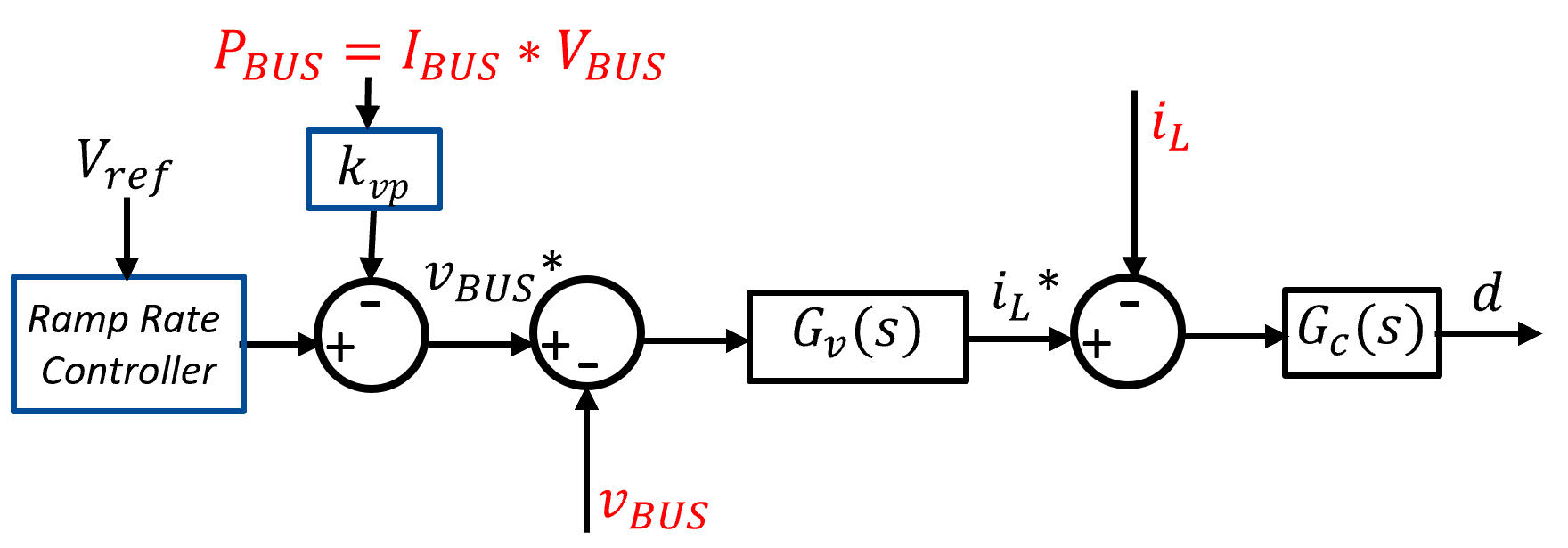}}
\caption{Finalized control scheme for CE+T equivalent model}
\label{fig}
\end{figure}
\section{Verification of Modeling}
Table 1 parameters are used  to implement the control scheme of Fig 7 for time domain equivalent model of COTS converter - implemented in MATLAB Simulink. For hardware characterization, CE+T is connected to Chroma 1\textit{000V/100A/5.2KW} electronic load. Line length between electronic load and CE+T is short and its impedance is ignored for this section. But additional \textit{760uH} line inductance is inserted between load and source converter to verify the instability, that is explained in Subection B of section IV. Results of time domain simulation in MATLAB Simulink vs hardware characterization of COTS converter are compared to verify the modeling approach. 
\subsection{Verification of Current Controller}\label{AA}
 For an approximate switching frequency of \textit{30KHz} (verified from CE+T) , CE+T's current control loop have a bandwidth of \textit{20KHz}. A step response of current controller from both time domain simulation and CE+T are shown in Fig. 8 and Fig. 9 respectively, verifying the modeling of current control loop. BW from step response is estimated by modeling the current control as first order system's time constant. 
\begin{figure}[htbp]
\centerline{\includegraphics[scale=0.45]{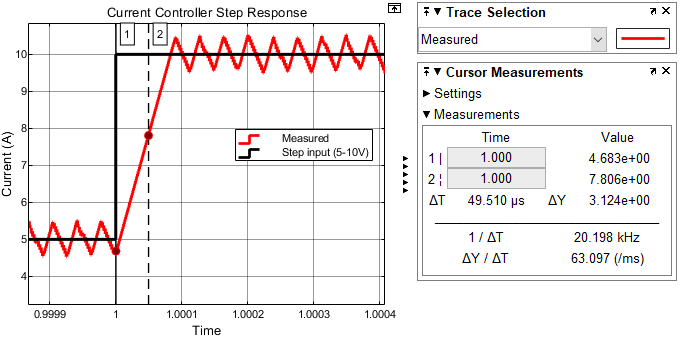}}
\caption{Current controller BW estimation from time domain model}
\label{fig}
\end{figure}
\begin{figure}[htbp]
\centerline{\includegraphics[scale=0.6]{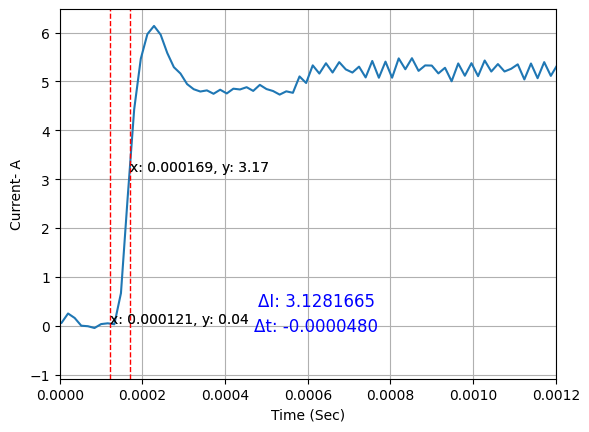}}
\caption{Current controller BW estimation from CE+T converter}
\label{fig}
\end{figure}

As we know, change in current for one time constant for \textit{5A} step is \textit{~3.2A} (\textit{63} percent of \textit{5A}), as shown for 5\textit{-10A} step in Fig 8 and \textit{0-5A} step in Fig 9. Estimated time for one time constant is \textit{0.00005sec} which corresponds to \textit{20KHz} BW of current controller. This section verifies that the current controller BW in time domain model is close to the CE+T current controller. 
\subsection{Verification of Voltage Controller and Droop Controller}\label{AA}
As discussed in subsection B and C in section II, CE+T follows the ramp rate controller for any change in voltage reference, and a Low pass filter like response in series with droop cofficient. Fig 10 shows the time domain simulations where initially at \textit{t=0sec},  converter is regulating the constant \textit{350Vdc} and supply \textit{3600W} CPL, and at t\textit{=1sec}, PV droop (\textit{Kvp=60V/3600W}) is enabled and converter voltage drops to \textit{290V} by following a designed \textit{200Hz} LPF response. Then voltage are restored by adjusting the voltage reference (at \textit{t=2sec}), that follows the ramp rate of \textit{50V/sec} - taking \textit{1.2secs} to restore \textit{60V} difference. 
\begin{figure}[htbp]
\centerline{\includegraphics[scale=0.53]{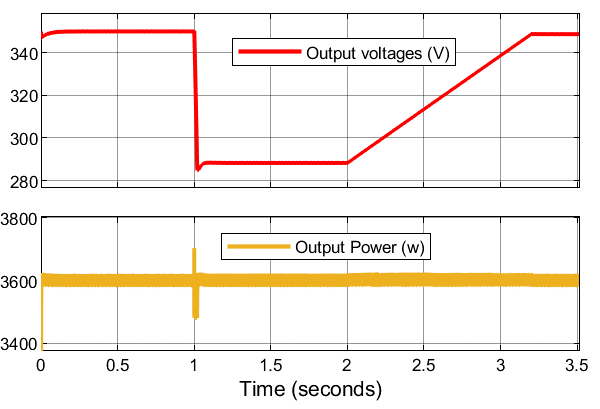}}
\caption{Voltage and droop loop characteristics from time domain model}
\label{fig}
\end{figure}
\begin{figure}[htbp]
\centerline{\includegraphics[scale=0.55]{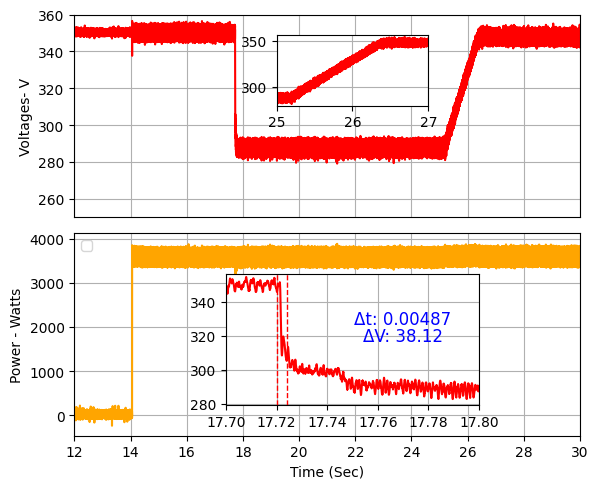}}
\caption{Voltage and droop loop characteristics from CE+T converter}
\label{fig}
\end{figure}
Fig 11 verifies the voltage ramp rate controller and droop characteristics of CE+T converter by putting the same parameters of droop coefficient,ramp rate and LPF in CE+T as in model. Load of \textit{3600W} is enabled at \textit{t=14sec} without droop, and then a droop of \textit{60V/3600W} is enabled at \textit{t=17.72sec} by following a \textit{200Hz} droop LPF. We can see in zoomed in snippet that the time constant of about \textit{0.00487sec} (corresponds to \textit{208Hz}) for droop BW. At the same time, voltage ramp rate of \textit{50V/sec} can be seen when voltage restoration command is given at\textit{ t=25.2sec}, restoring 60V difference in \textit{1.2sec}. This section verifies the mapping of steady state (droop coefficient) and dynamic (low pass filter) performance of equivalent model with COTS converter. 

\section{RESULTS: Use Cases of Equivalent Model}
This section proves the effectiveness of equivalent model by comparing the performance of the equivalent model to actual COTS converter for two different use cases. A) Performance under different droop coefficients, B) CPL instability in inductive DC microgrid. 
\subsection{Performance Under Different Droop Variable}\label{AA}
As the equivalent model uses VP droop, modulating the output voltage based on net power transfer to/from the converter. This case is performed under ideal transmission lines (line inductance and resistance are both zero) with constant power load connected at the output. Dynamics of droop (LPF bandwidth) is already verified in subsection B , section II of the article. This section verifies the steady state performance under different droop coefficients. Fig 12 (equivalent model) and Fig 13 (CE+T converter) shows the performance under two different droop gains \textit{10V/3600W} and \textit{20V/3600W} for a CPL load of \textit{3600W}. Two different droop gains results in two different steady state voltages; \textit{340V (Kvp=10V/3600W, K=1ohms)} and \textit{330V (Kvp=20V/3600W, K=2ohms),} can be seen in Fig 12 and Fig 13 correspondingly. 
\begin{figure}[htbp]
\centerline{\includegraphics[scale=0.65]{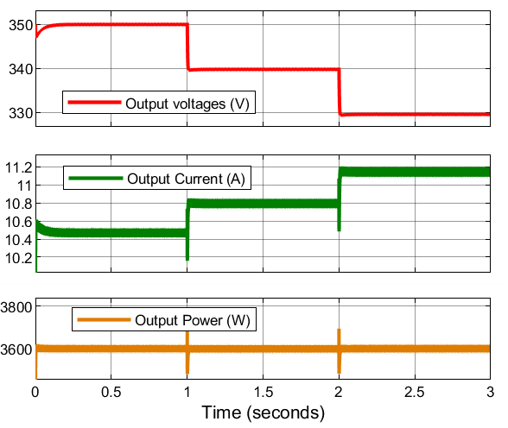}}
\caption{Varying droop gain results from time domain model}
\label{fig}
\end{figure}
\begin{figure}[htbp]
\centerline{\includegraphics[scale=0.7]{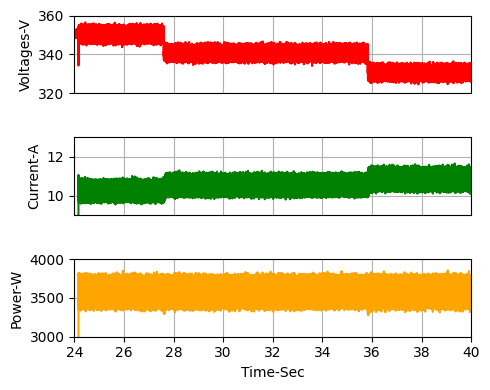}}
\caption{Varying droop gain results from CE+T converter}
\label{fig}
\end{figure}
\subsection{CPL Instability in DC Microgrids}\label{AA}
 Line inductance and bus capacitance together with CPL forms instability/oscillations in DC microgrids. This instability is modeled by different researchers [11][16,17]. As per [11], we need some minimum/base value of the bus capacitance \textit{(Co)} to stabilize a DC microgrid. 
 \begin{equation}
     (C=Co) > \frac{L}{KRe}
\end{equation}
Here \textit{L }is line inductance, \textit{K }is droop resistance, \textit{Re} is equivalent negative impedance of CPL. Oscillation/instability in CPL is because of the interaction of line inductance together with bus capacitance which limits the max CPL power transfer or minimum resistance for CPL power. Lets rearrange the above equation (6) as below to get (7). 
 \begin{equation}
     Lo < {KCRe}
\end{equation}
Lets assume we want to see the CPL instability for Bus capacitance \textit{C=30.8uF}, \textit{K=1 ohms,(Kvp=10V/3600W)} for a max power transfer of   \textit{4600W (Re=24.68 ohms)}, maximum inductance \textit{Lo} is considered to be \textit{760uH}. Lets evaluate the model performance against the mentioned parameters. As per Fig 14, converter was initially running without droop at \textit{4000W} and then load power start ramping at t=1 sec at ramping rate of \textit{1000W/sec}. We can see system goes unstable when CPL power reaches to about 4600W. Zoomed in snippets are shown for further clarity. Now lets evaluate the performance of CE+T under same set of parameters in Fig 15, where at t=65sec load of \textit{4000W }with droop \textit{(10V/3600W)} is enabled, then load is increased in \textit{100W steps} for each 5 seconds. We can see CE+T goes unstable at t=95sec and disconnects at around 97secs. Zoomed in snippets shows the oscillations in voltages, current and power. 
\begin{figure}[htbp]
\centerline{\includegraphics[scale=1]{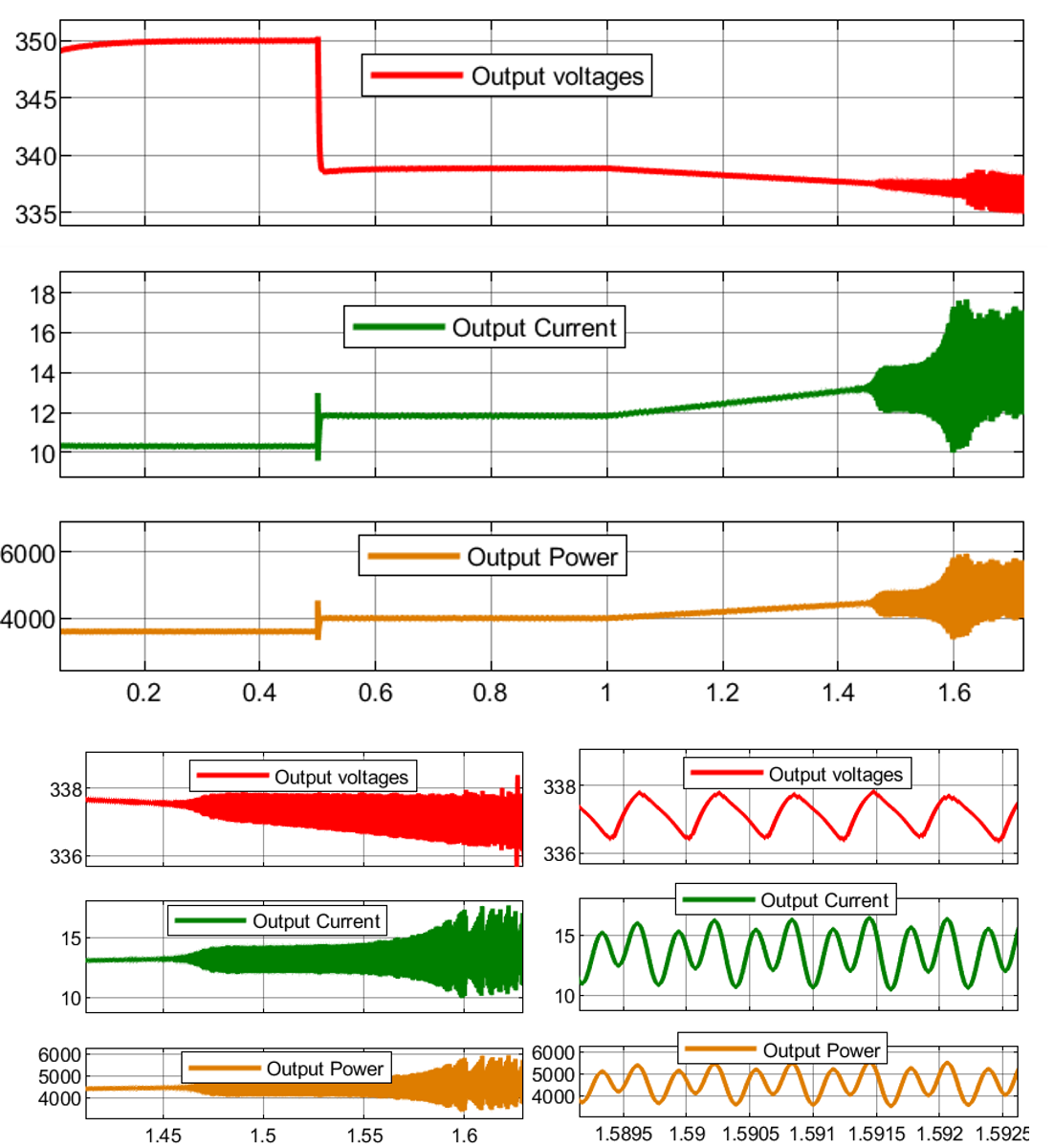}}
\caption{CPL instability from time domain model}
\label{fig}
\end{figure}
\begin{figure}
\centerline{\includegraphics[scale=0.85]{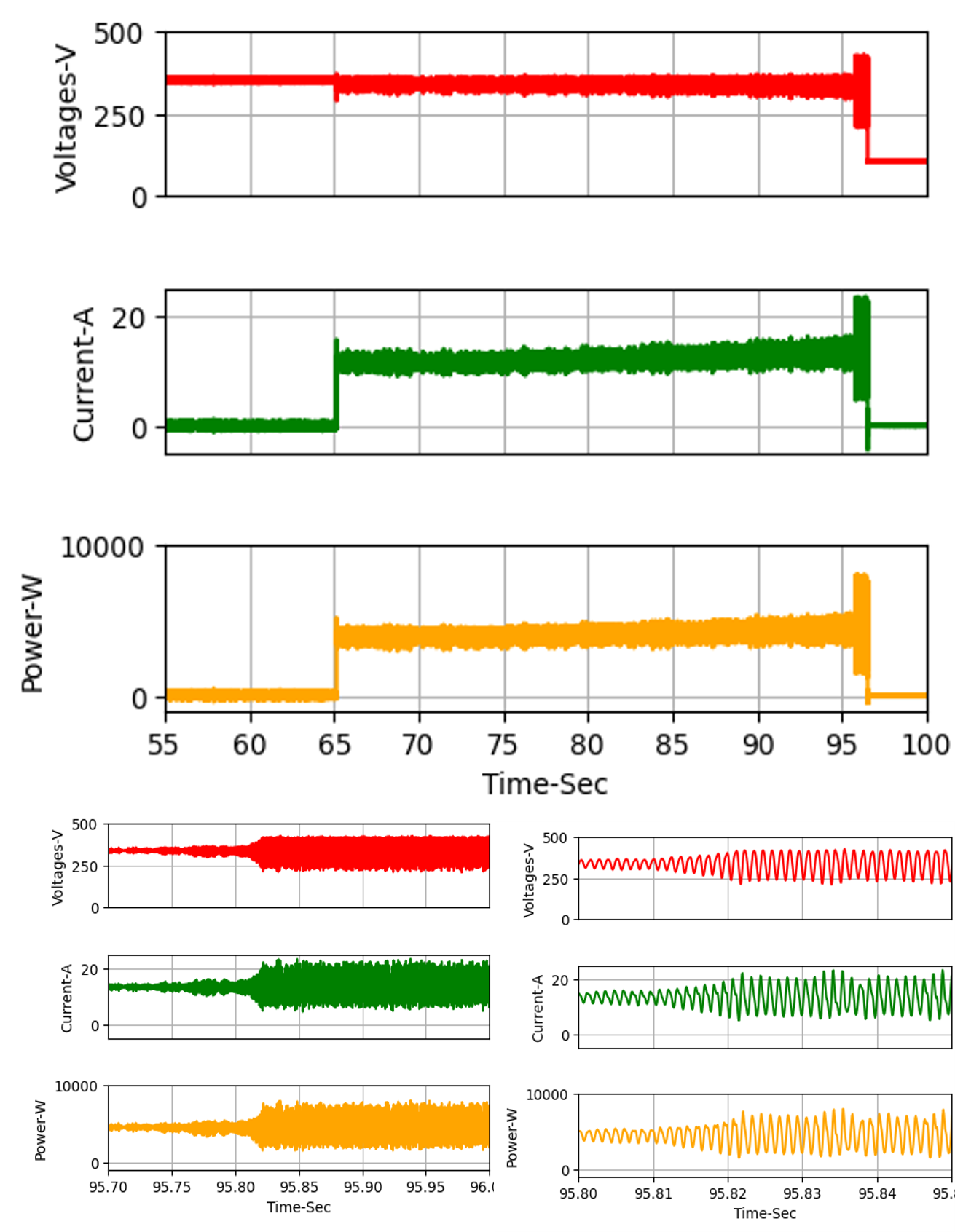}}
\caption{CPL instability from  CE+T converter}
\label{fig}
\end{figure}

This test case verifies that performance of equivalent model is close to actual converter, where power limit for CPL is same in both model and actual converter but the oscillation amplitude differs, this could be because of the different topology and different passives in the actual converter as compared to the model.

\section{Conclusion}
DC microgrids are a preferable choice over AC microgrids because of simplicity, easy control and direct integration of resources. Power electronics COTS products don't clearly reveal their topology and control structure and gains etc because of the IP and safety concerns. This article gives a baseline approach which can be used to build an equivalent model of any COTS converter to mimic the control performance. Proposed idea is evaluated on actual COTS CE+T Stabiliti 30C3 converter by comparing its performance with its equivalent model developed in MATLAB Simulink. Proposed approach models the control performance by assuming that the converter dynamics are faster than the control loops bandwidth. Keeping that in mind, proposed approach can be extended to  any type of the converter without knowing the topology and architecture. 

\end{document}